**$NbO_2$-based memristive neurons for burst-based perceptron**

*Yeheng Bo, Peng Zhang, Ziqing Luo, Shuai Li, Juan Song, and Xinjun Liu**

Y. Bo, P. Zhang, Prof. X. Liu
Tianjin Key Laboratory of Low Dimensional Materials Physics and Preparation Technology, Faculty of Science, Tianjin University, Tianjin 300354, People's Republic of China
E-mail: xinjun.liu@tju.edu.cn

Dr. S. Li
Unité Mixte de Physique, CNRS, Thales, Université Paris-Sud, Université Paris-Saclay, Palaiseau, France

Prof. J. Song
Department of Emergency Medicine, Putuo District Central Hospital, Shanghai University of Traditional Chinese Medicine, Shanghai 200062, People's Republic of China

Abstract—Neuromorphic computing using spike-based learning has broad prospects in reducing computing power. Memristive neurons composed with two locally active memristors have been used to mimic the dynamical behaviors of biological neurons. In this work, the dynamic operating conditions of $NbO_2$-based memristive neurons and their transformation boundaries between the spiking and the bursting are comprehensively investigated. Furthermore, the underlying mechanism of bursting is analyzed and the controllability of the number of spikes during each burst period is demonstrated. Finally, pattern classification and information transmitting in a perceptron neural network by using the number of spikes per bursting period to encode information is proposed. The results show a promising approach for the practical implementation of neuristor in spiking neural networks.

## 1. Introduction

Neuromorphic computing based on artificial neural network (ANN) has received extensive attention due to its low energy consumption. The high-power consumption in the conventional CMOS hardware based on the von Neumann framework limits the use for data-driven machine learning tasks. [1] From a software perspective, the energy consumed for training an ANN can be reduced by optimizing the algorithm. However, brain-inspired neuromorphic hardware has demonstrated promising potential for energy efficient computation. Replacing a portion of the CMOS components with emerging devices can implement an artificial neural network circuit using fewer components and lower power consumption than conventional CMOS. [2-3]

Artificial neuron is one of the most important elements in artificial neural networks. [4] Memristor is one of the promising candidates because of its low power consumption, plasticity and compatibility with conventional CMOS. [5, 6-7] Pickett et al. [8-9] demonstrated a memristive neuron by using two Mott memristors and realized the four basic neuronal functions, including: all-or-nothing spiking of an action potential, a bifurcation threshold to a continuous spiking regime, signal gain and a refractory period. Yi et al. [10] achieved 23 types of biological neuronal behaviors in memristive neurons, which possessed most of the known biological neuronal dynamics. Furthermore, Cassidy et al. [11] demonstrated the potential of achieving thousands of logic gates in neurons. Comparing to the traditional CMOS based artificial neurons, memristive neurons greatly reduce the power consumption and the number of components. However, there is still a lack of research on the state dynamics and operational window of the neuronal behaviors in relation to the input signals which limits the use of the rich dynamics of the memristive neuron. Comprehensive understanding of the memristive neurons is crucial for the practical implementation in artificial neural networks.

In this work, we report the spiking dynamics of $NbO_2$-based memristive devices that exhibit insulator–metal transition (IMT). The transformation conditions and operational boundary of $NbO_2$-based memristive neurons are investigated. The effect of input resistance and capacitance on the bursting behavior of the memristive neuron is studied. Furthermore, the memristive neurons are incorporated into a 9×1 array perceptron to demonstrate the potential for neuromorphic computing. Potential information spreading between neurons with different layers is also demonstrated.

## 2. Bursting behavior of memristive neuron

### 2.1. Circuit and model for memristive neuron

**Figure 1**a shows a biological neuron that generates an action potential in the direction of an output synapse after receiving sufficient stimulus from dendrites. **Figure 1**b shows the circuit diagram of $NbO_2$-based memristive neuron. The ion ($K^+$ or $Na^+$) channel consists of a $NbO_2$-based memristor and an opposite voltage source. Capacitors $C_1$ and $C_2$ in parallel with the ion channel are membrane capacitors. The resistor $R_2$ couples the two channels together with the input resistor to form a $NbO_2$-based memristive neuron. Memristive neuron dynamics of the output voltage $V_{out}$ will change subject to different input signals by adjusting the input resistance $R_{in}$ and input voltage $V_{in}$.

Lim et al.[12] approximated the resistance of the memristor to a hard switching between two preset resistance values of $R_{on}$ and $R_{off}$, and calculated the boundaries A and B of the memristive neuron that can generate spike when one of the memristors $X_1$ or $X_2$ is in the critical state of switching, as shown in **Figure 1**c,d. However, memristor is a non-linear resistor, and the memristor resistance at the critical state is different from $R_{on}$ and $R_{off}$, which results in inaccurate theoretical boundaries.

A resistor $R_{th}$ at the threshold voltage and a resistor $R_h$ at the hold voltage are introduced, as shown in **Figure 1**e. **Figure 1**f shows the operational window $R_{in}$-$V_{in}$ of the simulated (see the parameters used for simulation and calculation are shown in Table **S1** and **S2** in the supplementary material 1). The operational window diagram is divided into three main areas: failure to fire (white), continuous spike (blue), bursting spike (green). The theoretical boundaries after the introduction of $R_h$ and $R_{th}$ are optimized from **A** (or **B**) to **A'** (or **B'**), which are consistent with the simulated window diagram (see Equations S3, S4, S7 and S8 in the supplementary material 2).

Interestingly, a new boundary **C'** for the two dynamic transformations between continuous spike and bursting spike is theoretically introduced considering that $X_1$ and $X_2$ are simultaneously in the critical state of transition (see Equations S5 and S6 in the supplementary material 2), and the calculated boundary is consistent with the simulated boundary. According to the calculation formula of the three boundary lines, we can design the desired window size of the continuous spike and the bursting spike.

## 2.2. Controllable bursting behaviors under the DC input voltage

Generally, every bursting spike possesses two oscillations components: one is the inter-spike oscillation which is a fast spiking oscillation within a single burst, the other is inter-burst oscillation which is modulated by a slow oscillation between the bursts (see Figure S2a in the supplementary material 3).[13] In memristive neuron, $V_K$ represents fast spiking oscillation and $V_{Na}$ represents slow oscillation. **Figure 2**a-e are the waveforms of $V_{Na}$ and $V_K$ when the capacitance $C_1$ is in the range between 2 nF and 5 nF and $C_2$ is between 0.2 nF and 0.5 nF. **Equation 1** and **2** describe the charging and discharging speed of $V_{Na}$ and $V_K$, respectively:

$$\frac{dV_{Na}}{dt} = \frac{1}{C_1}\left(\frac{V_{in}-V_{Na}}{R_{in}} + \frac{V_K-V_{Na}}{R_2} - \frac{V_{Na}-V_1}{R_{X1}}\right) \quad (1)$$

$$\frac{dV_K}{dt} = \frac{1}{C_2}\left(\frac{V_{Na}-V_K}{R_2} - \frac{V_K-V_2}{R_{X2}}\right) \quad (2)$$

The increase of $C_2$ will cause the charging and discharging speed of $V_K$ to decrease, which will make the oscillation period of $V_K$ longer, as shown in **Figure 2**a-c. The increase of $C_1$ will cause the charging and discharging speed of $V_{Na}$ to decrease, which will make the oscillation period of $V_{Na}$ longer, as shown in **Figure 2**c-e. The mismatch between the charging and discharging speed of $V_{Na}$ and $V_K$ produces bursting spikes in biological membranes.

Practically, $C_1$ defines the spiking cycle (slow process) and $C_2$ defines the spike events within the cycle (fast process). By using a large $C_1$ and a small $C_2$ as well as opposite potentials $V_1$ and $V_2$, the slow-inward $Na^+$ and fast-outward $K^+$ ionic currents through the membrane are emulated, which is a sort of "transient engineering". **Figure 2**f shows the relationship between the $C_1/C_2$ ratio and the number of spikes during each burst period. As expected, the number of spikes/per cycle depends on the $C_1/C_2$ ratio. The rise in $C_1/C_2$ will lead to an increase in the number of spikes during each burst period. This implies that as the "$Na^+$ process" becomes slower in respect to the "$K^+$ process", it can accommodate more spikes. Furthermore, the charging and discharging speed of $V_{Na}$ and $V_K$ can be regulated not only by $C_1$ and $C_2$, but it can also be regulated by other parameters. Therefore, $V_{in}$, $V_1, V_2$, $R_{in}, R_2, R_{X1}$ and $R_{X2}$ in equations (1) and (2) can also be used to control the number of spikes.

### 2.3. Adjustable spiking numbers in the $R_{in}$-$V_{in}$ phase diagram

The number of spikes per period can be regulated by adjusting the $V_{in}$ and $R_{in}$. To demonstrate the bursting spike behaviors in details, the $R_{in}$-$V_{in}$ phase diagram is plotted with $C_1$=5 nF and $C_2$=0.5 nF, as shown in **Figure 3**a. The phase diagram is divided into three main areas: failure to fire (white), continuous spike (blue), bursting spike (other colors). In the bursting spike area, two kinds of spikes are observed with different conditions of $V_{in}$ and $R_{in}$, i.e. three-spike (olive) and four-spike (pink) responses during each burst period.

When $C_1$ is increased to 10 nF with a fixed $C_2$ of 0.5 nF, the similar three-area $R_{in}$-$V_{in}$ phase diagram is obtained, as shown in **Figure 3**b. In the bursting spike area, there are five kinds of spikes from five-spike (green) to nine-spike (red) responses during each burst period.

**Figure 3**c shows the the effect of $C_1$ on the maximum number range of spikes in each burst period, where the $V_{in}$ is fixed at 400 mV and $R_{in}$ varies within the proper range. When $C_1$ is increased from 5 nF to 25 nF, the number range of spikes in each burst period increases from (3-4) to (14-22). Nine

kinds of spikes can be obtained when $C_1$ = 25 nF. Larger the number range of spikes can carry more information, which is very important in pattern classification and neuromorphic computing. [14-15]

## 3. Exploring potential application for bursting spike

### 3.1. Burst-based perceptron for pattern classification

A perceptron is an algorithm that produces an output by applying the weighted sum of the input values through an activation function. The activation function in ANNs loosely represents the firing rate of biological neurons, where there is a nonlinear relationship between the firing rate and the input. Here, the nonlinear activation function of the memristive neuron show that the number of spikes per period decreases when the input current amplitude is increasing (see Figure S3 in the supplementary material 4).

In order to facilitate the construction of the burst-based perceptron, the memristive neuron circuit is symbolically represented as '**N**' element, as shown in **Figure 4**a. We used resistors as input synapses and memristive neurons as neurons to build a 9 × 1 array, as shown in **Figure 4**b. The simulated parameters of the memristive neurons are taken from Table **S1** except for $C_1$ = 25 nF (see in the supplementary material 1). To prove that bursting behaviour can be used for pattern classification, synaptic resistances are programmed with different weights. Three letter patterns, 'n', 'z' and 'v' are used as inputs as shown in **Figure 4**c. Nine squares correspond to the inputs of each synapse. A gray square means a 0.6 V input and a white square means a 0.1 V input. The number of spikes in each burst period corresponds to the three different modes 'n', 'z', and 'v' is 16, 17, 18. This shows that the bursting behavior of memristive neurons has the potential for pattern classification. The number of spikes during each burst period is used to encode information which is different from the conventional approach by using oscillation phase and frequency. [16-17] In addition, there are two other types of information embedded within each bursting which includes the inter-burst frequency and the duty cycle. Both of these features can be programmed as a function of input voltage (see Figure S2b and S2c in the supplementary material 3).

### 3.2. Information transition by bursting spike

The burst-based perceptron is potential to make the multi-layer neuron network. The information can be transformed between the adjacent layers. To prove that the bursting behavior of the memristive neurons can be transmitted from the upper neuron to the lower neuron, three memristive neurons are connected in series with two resistors as shown in **Figure 5**a. When $C_1$ = 10 nF and $C_2$ =1 nF in neuron $N_1$, $C_1$ = 3 nF and $C_2$ =0.3 nF in neuron $N_2$, $C_1$ = 1 nF and $C_2$ =0.08 nF in neuron $N_3$, $R_m$ =10 kΩ, $R_n$ =10 kΩ and $R_q$ =15 kΩ, bursting behavior can stably propagate from the neuron in the previous layer to the neuron in the next layer, as shown in **Figure 5**b. Neuron $N_1$ generates bursting signals at the input of a DC voltage source. Subsequently, a bursting signal with the number of 5 spikes in each burst period generated by neuron $N_1$ will excite the next neuron $N_2$, and neuron $N_2$ will generate the same number of spikes in each burst period after a short delay after $N_1$ excitation. Compared with the bursting of neuron $N_1$, the number of spikes in each burst period of neuron $N_2$ is the same, but inter-spike period of neuron $N_2$ is shorter. Similarly, neuron $N_3$ can receive the bursting signal generated by neuron $N_2$, thereby generating a bursting signal of the same number of spikes in each burst period but with a shorter inter-spike period. When $C_1$ = 8 nF and $C_2$ =4 nF in neuron $N_1$, $C_1$ = 3.5 nF and $C_2$ =2 nF in neuron $N_2$, $C_1$ = 2 nF and $C_2$ =1 nF in neuron $N_3$, $R_m$ =10 kΩ, $R_n$ =10 kΩ and $R_q$ =15 kΩ, a single spike can also spread between neurons, as shown in **Figure 5**c. This shows that the bursting signal can keep the number of spikes in each burst period transmitted from the memristive neuron in the previous layer to the memristive neuron in the next layer.

## 4. Conclusion

In conclusion, we investigated the rich dynamics and the transition conditions between different dynamic behaviors of $NbO_2$-based memristive neuron. A memristor model is used to calculate three theoretical boundaries of the memristive neuron for transformation of the dynamics. The three boundaries are consistent with the simulation. More specifically, $NbO_2$ devices with dynamics in different timescales are coupled in order to emulate the slow and fast dynamics of $Na^+$ and $K^+$ ionic channels in biological membranes during the creation of action potentials. The impact of $C_1$ and $C_2$ on the bursting behaviors was explored. The rise in $C_1/C_2$ will lead to an increase in the number of spikes in each burst period. When $C_1$ becomes larger, the number range of the spikes per period increases, then the bursting can carry more information. A burst-based perceptron for pattern classification by using memristive neurons was proposed. Bursting behaviors are shown to propagate the signals from the upper memristive neuron to the lower memristive neuron. The results in this work propose a promising approach to implement the rich dynamics of the memristive neurons for neuromorphic computing.

**Supporting Information**
Supporting Information is available from the Wiley Online Library or from the author.

**Acknowledgements**
This work was potential supported by the future granted funding.

Received: ((will be filled in by the editorial staff))
Revised: ((will be filled in by the editorial staff))
Published online: ((will be filled in by the editorial staff))

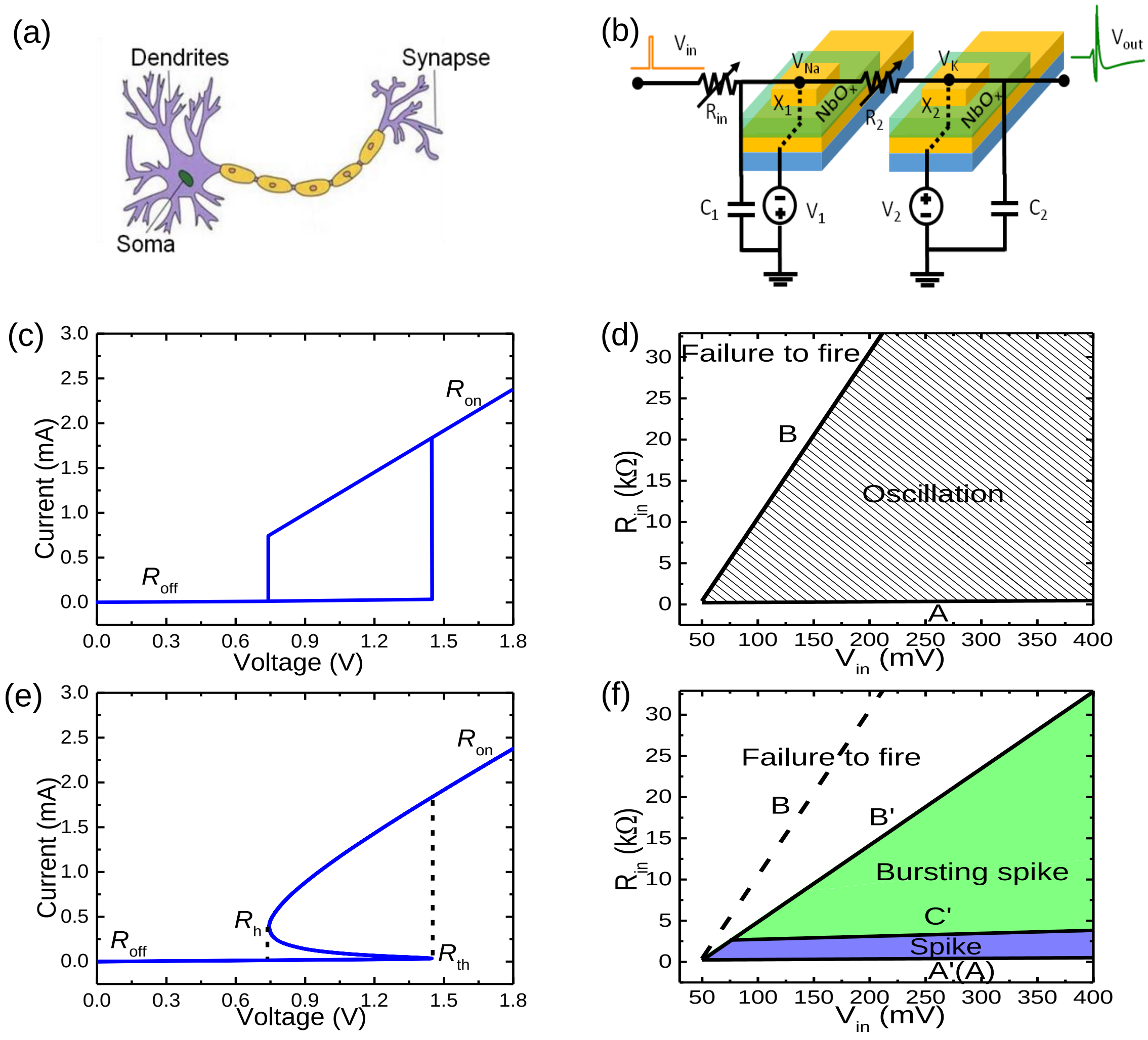

**Figure 1.** (a) Schematic structure of a biological neuron. (b) Circuit diagram of memristive neuron composed by two $NbO_2$ active memristors. (c) Current-Voltage (*I*-*V*) curve simulated by the simplest model with fixed values of $R_{on}$ and $R_{off}$ of the memristor and (d) the corresponding operation window diagram calculated by memristive neurons circuit. The two boundaries A and B marked by solid lines divide the window into two areas: failure to fire (white), oscillation (shaded). (e) *I*-*V* curve simulated by the IMT model of memristor and (f) the corresponding operation window diagram calculated by memristive neurons circuit. The three boundaries A', B', and C' marked by solid lines divide the window into three areas: failure to fire (white), continuous spike (blue), bursting spike (green). The simulation results of the optimized calculation verified the three boundaries, which is close to the dynamic behavior of the real memristive neurons system. Note that the dashed line is the boundary B marked in (d).

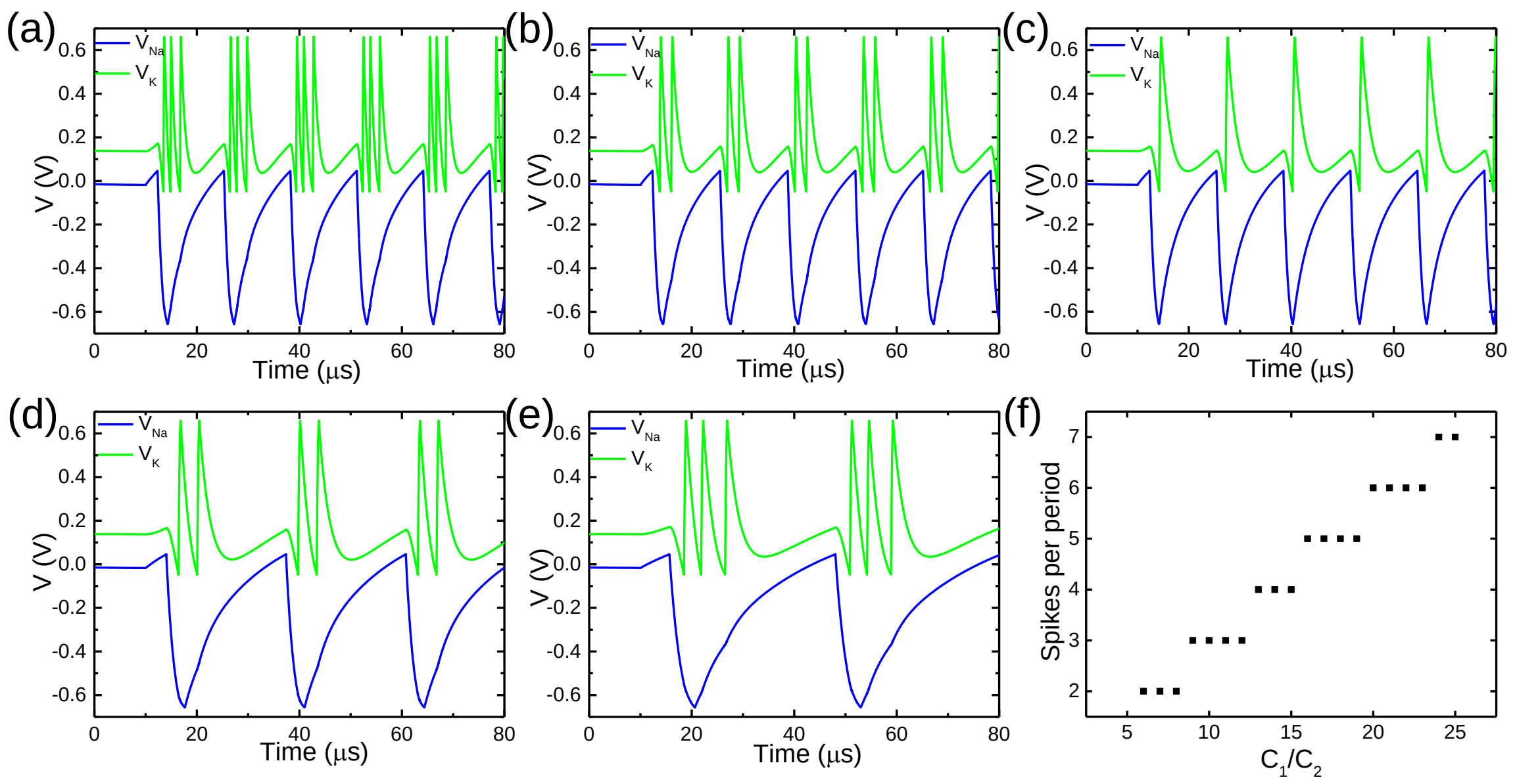

**Figure 2.** (a-e) $K^+$ and $Na^+$ ion channel voltage waveforms. (a) $C_1$=2 nF, $C_2$=0.2 nF. (b) $C_1$=2 nF, $C_2$=0.3 nF. (c) $C_1$=2 nF, $C_2$=0.5 nF. (d) $C_1$=3.5 nF, $C_2$=0.5 nF. (e) $C_1$=5 nF, $C_2$=0.5 nF. (f) Effect of $C_1/C_2$ on the number of spikes in each burst period.

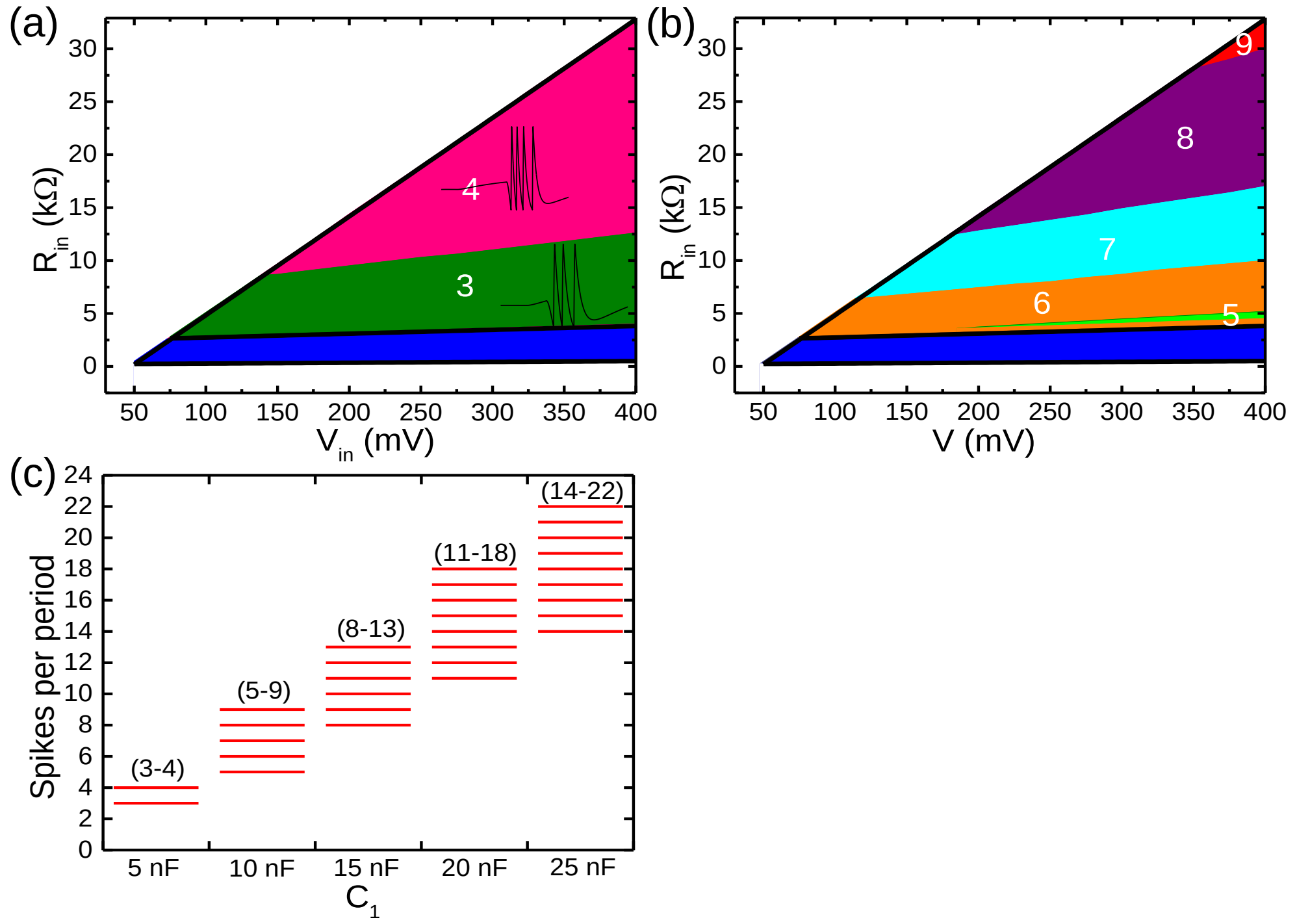

**Figure 3.** The simulated operation window of memristive neuron in $R_{in}$-$V_{in}$ plots, where the used parameters are (a) $C_1$=5 nF, $C_2$=0.5 nF and (b) $C_1$=10 nF, $C_2$=0.5 nF, respectively. The bursting area is divided into some sub-areas according to the number of spikes in each burst period. The insets of (a) show the waveform illustrations of 3 spikes (olive area) and 4 spikes (pink area) per period. (c) The effect of $C_1$ on the number range of spikes per period.

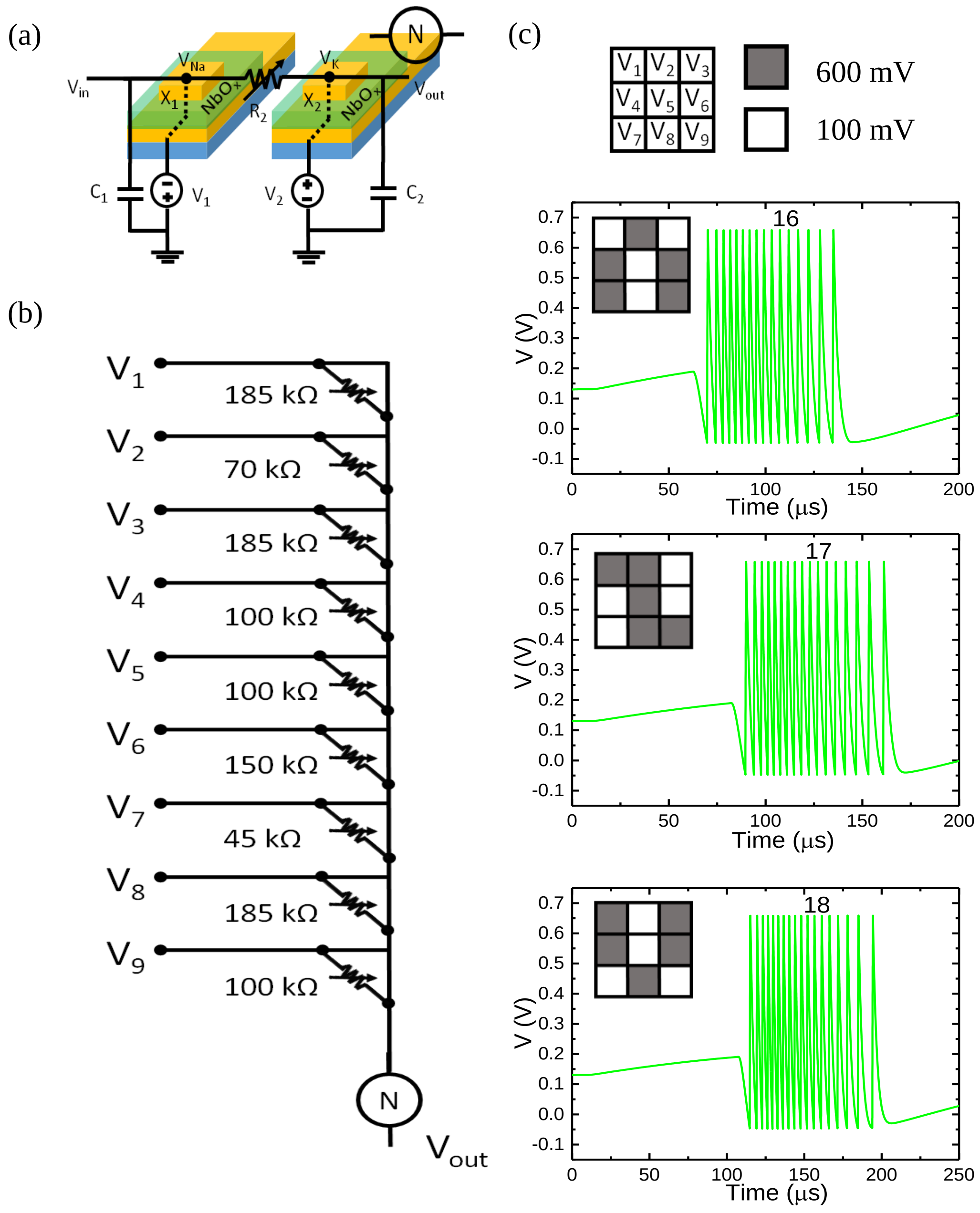

**Figure 4.** (a) Memristive neuron symbol and its internal circuit diagram. (b) Schematic of burst-based perceptron: 9×1 array using a memristive neuron. (c) The number of spikes in each burst period of memristive neurons as a response to the 'n', 'z' and 'v' input patterns.

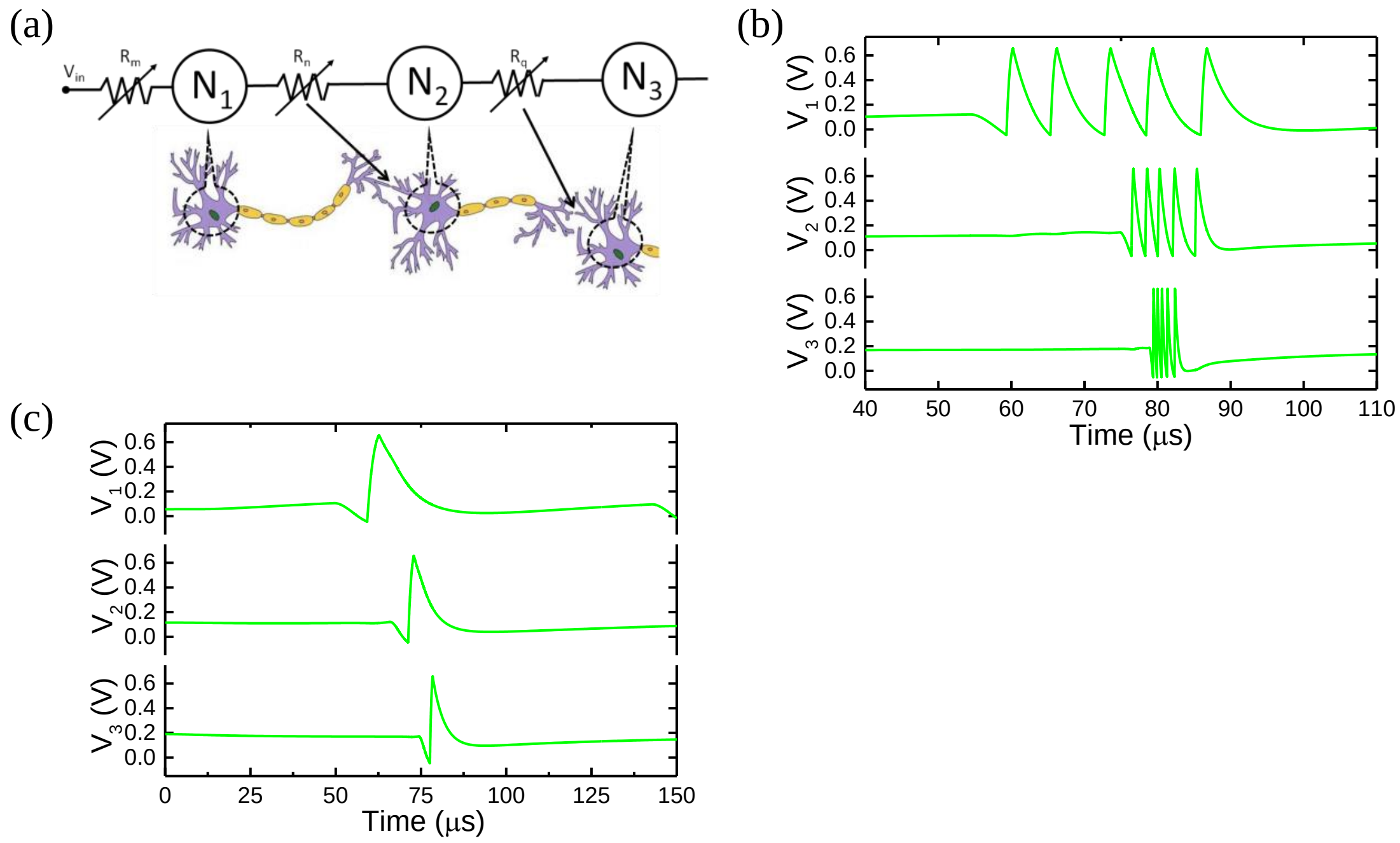

**Figure 5.** (a) Schematic diagram of biological neurons and memristive neurons. (b-c) Output signals of neuron $N_1$, $N_2$ and $N_3$.

# Supporting Information

## $NbO_2$-based memristive neurons for burst-based perceptron

*Yeheng Bo, Peng Zhang, Ziqing Luo, Shuai Li, Juan Song, and Xinjun Liu**

### 1) Parameters for circuits and memristor model used in the simulation

**Table S1.** Parameters used for calculation.

| Parameter | Model value (SI) | Parameter | Model value (SI) |
|---|---|---|---|
| $C_1$ | 5 nF | $C_2$ | 0.5 nF |
| $V_1$ | -1.4 V | $V_2$ | 1.4 V |
| $R_{th}$ | 41 kΩ | $R_h$ | 1.98 kΩ |
| $R_{off}$ | 49 kΩ | $R_{on}$ | 0.85 kΩ |
| $V_h$ | 746 mV | $V_{th}$ | 1448 mV |
| $R_2$ | 6 kΩ | $R_{in}$ | 0-33 kΩ |

**Table S2.** Memristor parameters used in the simulation. [R1]

| Parameter | Model value (SI) | |
|---|---|---|
| $T_{IMT}$ | 1080 K | Insulator-mental transition temperature |
| $T_{amb}$ | 296 K | Ambient temperature |
| $\kappa$ | 1.5 W·m$^{-1}$·K$^{-1}$ | Thermal conductivity |
| $\hat{c}^p$ | 2.6×10$^{6}$ J·m$^{-3}$·K$^{-1}$ | Volumetric heat capacity |
| $\rho_{met}$ | 1×10$^{-4}$ Ω·m | Metallic phase electrical resistivity |
| $\rho_{ins}$ | 7×10$^{-3}$ Ω·m | Insulating phase electrical resistivity |
| $r_{ch}$ | 30 nm | Conduction channel radius |
| $L$ | 20 nm | Conduction channel length |
| $\Delta\hat{h}^{tr}$ | 1.6×10$^{8}$ J·m$^{-3}$ | Volumetric enthalpy of transformation |

### 2) Calculation for three main boundaries in $R_{in}$-$V_{in}$ plots for operational window

**Figure S1**a shows the simulated operational window in $R_{in}$-$V_{in}$ plots. In order to assist in the study, we select the parameters of the six points $\alpha^-$($R_{in}$=0.504 kΩ), $\alpha^+$($R_{in}$ =0.505 kΩ), $\beta^-$($R_{in}$ =3.79 kΩ), $\beta^+$($R_{in}$ =3.8 kΩ), $\gamma^-$($R_{in}$ =32.9 kΩ) and $\gamma^+$($R_{in}$ =33 kΩ) which are located near the upper and lower sides of the three boundary line **A'**, **B'** and **C'** of **Figure S1**a. The spiking dynamics is described by the membrane potential $V_{Na}$ and $V_K$ as follows: [R2]

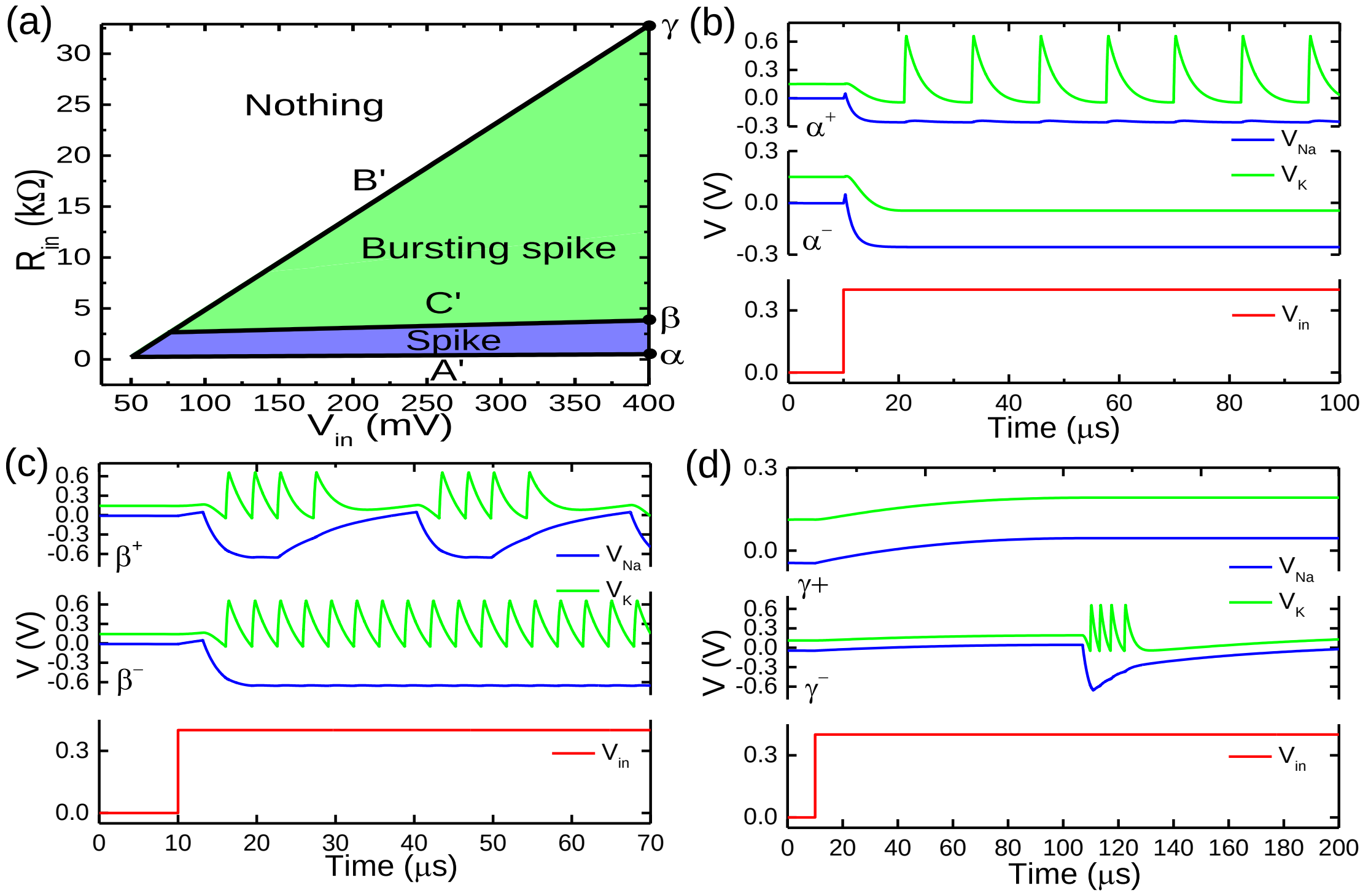

**Figure S1.** (a) Simulated and calculated window of $R_{\mathrm{in}}$-$V_{\mathrm{in}}$. The blue area and the green area are simulated window diagrams. The three lines A', B' and C' are theoretically calculated boundary lines. (b-d) Waveforms of $V_{\mathrm{K}}$ and $V_{\mathrm{Na}}$ near three points $\alpha$, $\beta$, and $\gamma$. Where $\alpha^-$($R_{\mathrm{in}}$=0.504 kΩ), $\alpha^+$($R_{\mathrm{in}}$ =0.505 kΩ), $\beta^-$($R_{\mathrm{in}}$ =3.79 kΩ), $\beta^+$($R_{\mathrm{in}}$ =3.8 kΩ), $\gamma^-$($R_{\mathrm{in}}$ =32.9 kΩ) and $\gamma^+$($R_{\mathrm{in}}$ =33 kΩ).

$$C_1\frac{dV_{Na}}{dt}+\frac{V_{Na}-V_1}{R_{X1}}=\frac{V_{in}-V_{Na}}{R_{in}}+\frac{V_K-V_{Na}}{R_2} \tag{S1}$$

$$C_2\frac{dV_K}{dt}+\frac{V_K-V_2}{R_{X2}}=\frac{V_{Na}-V_K}{R_2} \tag{S2}$$

The $V_{\mathrm{Na}}$ and $V_{\mathrm{K}}$ waveform of the point $\alpha^-$ and $\alpha^+$ is as shown in **Figure S1**b. Before transitioning from $\alpha^-$ to $\alpha^+$, the threshold switch $X_1$ is in a metallic state and the threshold switch $X_2$ is in a critical state in which the insulating state transitions to a metallic state, $V_2-V_K=V_{th}$. It can be seen from $\alpha$ in **Figure S1**b that $\frac{dV_{Na}}{dt}\approx 0$ and $\frac{dV_K}{dt}\approx 0$ after the circuit is stable. Bringing the limit condition back to equations (1) and (2), the equation at the boundary line **A'** is as follows:

$$R_{in}=mV_{in}+m\left(\frac{V_{th}(R_{X2}+R_2)}{R_{X2}}-V_2\right) \tag{S3}$$

$$m=\frac{R_{X1}R_{X2}}{R_{X2}(V_2-V_1-V_{th})-V_{th}(R_2+R_{X1})} \tag{S4}$$

for case of $R_{X1}=R_{\mathrm{on}}$ and $R_{X2}=R_{\mathrm{th}}$.

The $V_{Na}$ and $V_K$ waveform of the point $\beta^-$ and $\beta^+$ is as shown in **Figure S1**c. When transitioning from $\beta^-$ to $\beta^+$, bursting activity correspond to an infinite spike train interrupted by short periods of quiescence. The threshold switch $X_1$ is in a critical state in which the metallic state transitions to an insulating state, the threshold switch $X_2$ is in a critical state in which the insulating state transitions to a metallic state, $V_2 - V_K = V_{th}$, $V_{Na} - V_1 = V_h$. At the transition point of the graph $\delta$, $\frac{dV_{Na}}{dt} \approx 0$. Bringing the limit condition back to equations (1) and (2), the equation at the boundary line **B'** is as follows:

$$R_{in} = nV_{in} + n(V_h + V_1) \quad \text{(S5)}$$

$$n = \frac{R_{X1}R_2}{R_2V_h - R_{X1}(V_2 - V_{th} - V_h - V_1)} \quad \text{(S6)}$$

for case of $R_{X1} = R_h$.

**Figure S1**d shows the $V_{Na}$ and $V_K$ waveforms of point $\gamma^-$ and $\gamma^+$. When transitioning from $\gamma^+$ to $\gamma^-$, bursting activity correspond to an infinite period of quiescence interrupted by groups of spikes. The threshold switch $X_1$ is in a critical state in which the insulating state transitions to a metallic state, and the threshold switch $X_2$ is in an insulating state, $V_{Na} - V_1 = V_{th}$. It can be seen from $\gamma^+$ in **Figure S1**d that $\frac{dV_{Na}}{dt} \approx 0$ and $\frac{dV_K}{dt} \approx 0$ after the circuit is stable. Bringing the limit condition back to equations (1) and (2), the equation at the boundary line **C'** is as follows:

$$R_{in} = qV_{in} + q(V_{th} + V_1) \quad \text{(S7)}$$

$$q = \frac{R_{X1}(R_2 + R_{X2})}{V_{th}(R_2 + R_{X2}) - R_{X1}(V_2 - V_{th} - V_1)} \quad \text{(S8)}$$

for case of $R_{X1} = R_{th}$ and $R_{X2} = R_{off}$, respectively.

**3) Other dynamic characteristics of bursting signals**

Bursting provide two important information: inter-burst frequency and duty cycle. The inter-burst period includes two parts as the active phase and quiescent period, as shown in **Figure S2**a. The duty cycle is defined as

$$duty\ cycle = \frac{T_{\text{active phase}}}{T_{\text{inter-burst period}}} \quad \text{(S9)}$$

Using the circuit parameters in Table **S1**, $R_{in}$ = 6 kΩ, 8 kΩ, 13 kΩ, and 15 kΩ are selected in 3 spikes (lines) and 4 spikes (scatters) bursting areas to obtain the relationship between $V_{in}$ and duty cycle, as shown in **Figure S2**b. The duty cycle increases with increasing $V_{in}$ in the same spikes area. When the input voltage is fixed, the larger the input resistance, the smaller the duty cycle.

$R_{in}$ = 6 kΩ, 8 kΩ, 13 kΩ, and 18 kΩ are selected in 3 spikes (line) and 4 spikes (point) bursting areas to obtain the relationship between $V_{in}$ and inter-burst frequency, as shown in **Figure S2**c. The inter-burst frequency increases as the $V_{in}$ increases. The duty cycle and inter-burst frequency can change as the input changes, which means that information can be encoded into the duty cycle and inter-burst frequency.

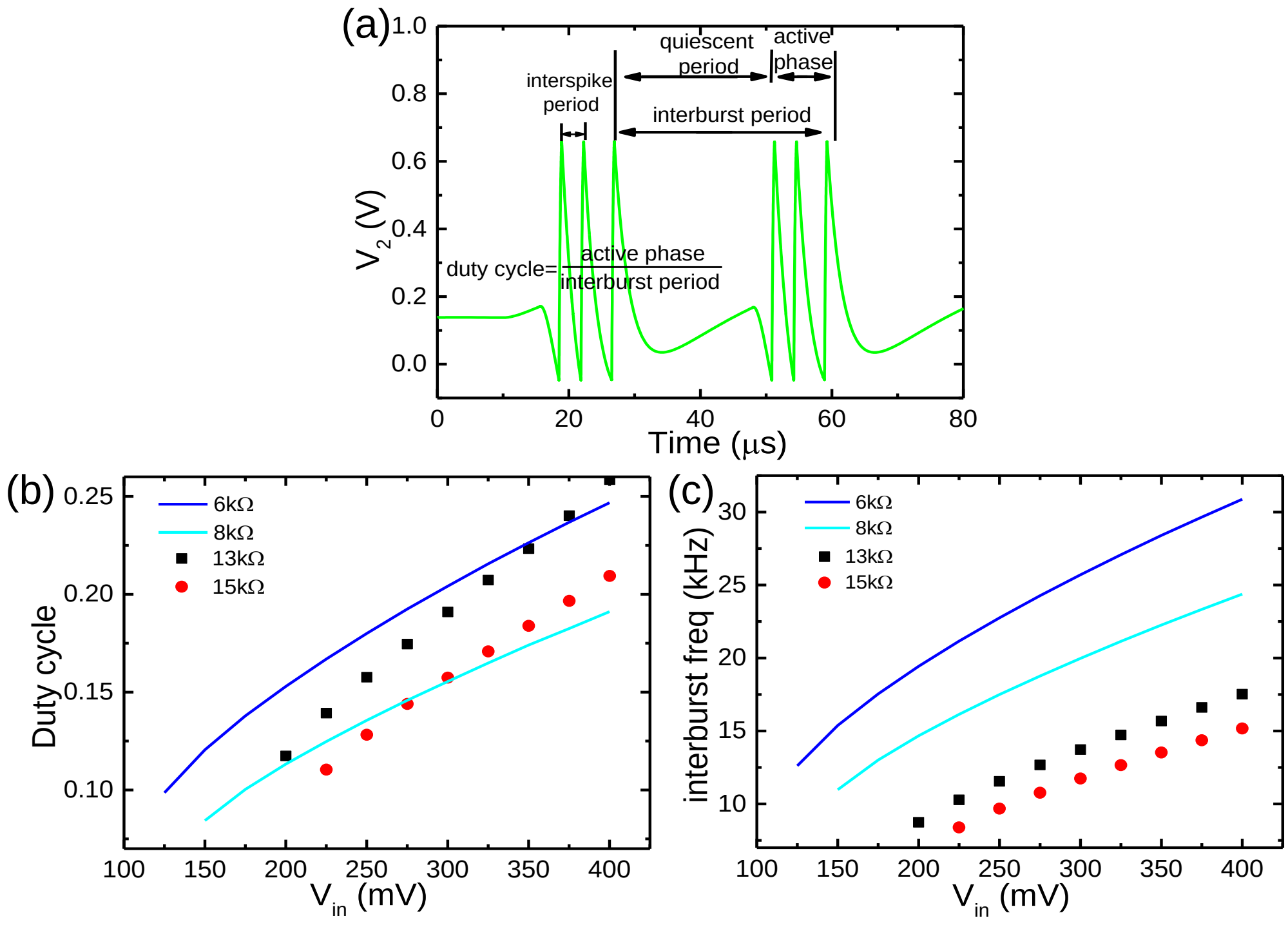

**Figure S2.** (a) The bursting spikes of the memristive neuron for defining the new parameter of duty cycle, which is the ratio between the active phase and the inter-burst period. (b) Duty cycle and (c) the inter-burst frequency as a function of $V_{in}$, respectively. Both the inter-burst frequency and duty cycle of bursting can be designed as the neuronal information carriers.

### 4) The activation function of the burst-based perceptron

The main function of the perceptron is to transform the input signal into an output signal through an activation function. [R3] The number of spikes per period decreased as the input current amplitude increased, as shown in Figure S3.

For the application in the spiking neural networks, it is important to show how an activation function (relationship between input current and spiking numbers) can be constructed with the proposed devices. Here, the activation function of the burst-based perceptron can be used as the Exponential Units (EU) function, as plotted in Figure S3:

$$\mathbf{F(x)} = \begin{cases} \mathbf{0}, & \mathbf{if\ x} < m \\ \mathbf{a + be^{-kx}}, & \mathbf{otherwise} \end{cases} \quad \text{(S10)}$$

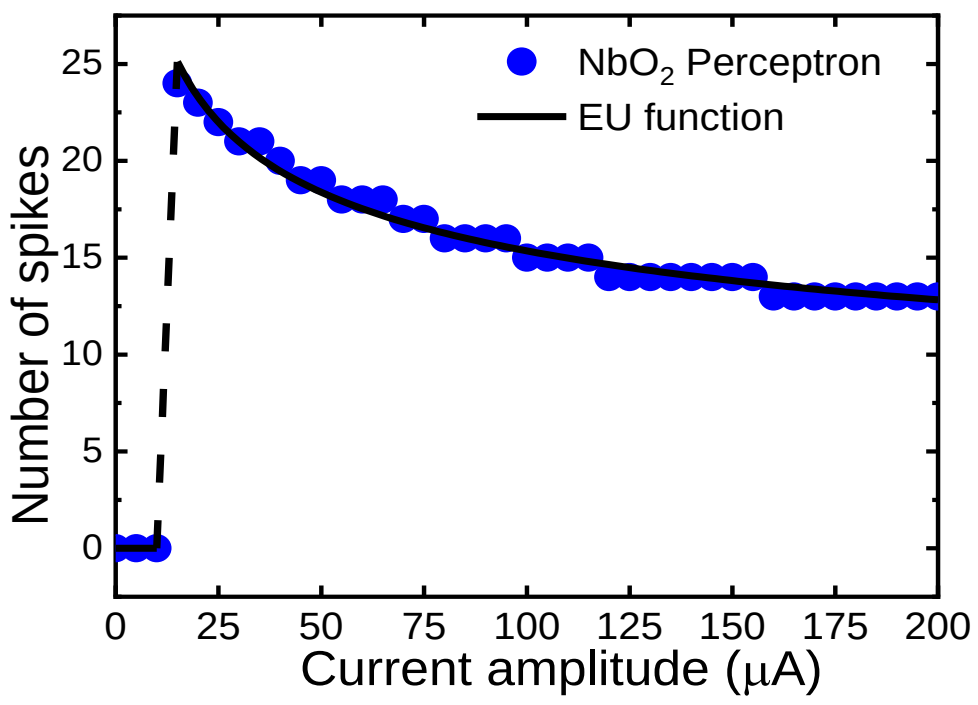

**Figure S3.** Number of spikes versus input current amplitude, where the used circuit parameters are $C_1$=25 nF, $C_2$=0.5 nF.